\documentclass{article}

\usepackage{PRIMEarxiv}

\usepackage[utf8]{inputenc} 
\usepackage[T1]{fontenc}    
\usepackage{hyperref}       
\usepackage{url}            
\usepackage{booktabs}       
\usepackage{amsfonts}       
\usepackage{nicefrac}       
\usepackage{microtype}      
\usepackage{lipsum}
\usepackage{fancyhdr}       
\usepackage{graphicx}       
\graphicspath{{media/}}     

\usepackage{subcaption}

\title{Efficient and accurate defect level modelling in monolayer MoS$_2$ via
GW+DFT with open boundary conditions
}

\author{
  Guido Gandus, Youseung Lee, Leonard Deuschle, Mathieu Luisier \\
  Integrated Systems Laboratory \\
  ETH  \\
  Z\"urich, Switzerland \\
   \And
  Daniele Passerone \\
  nanotech@surfaces \\
  EMPA \\
  Z\"urich, Switzerland\\
}

\begin{document}
\maketitle

\begin{abstract}
Within the framework of many-body perturbation theory (MBPT) integrated with density functional theory (DFT), a novel defect-subspace projection GW method, the so-called p-GW, is proposed. By avoiding the periodic defect interference through open boundary self-energies, we show that the p-GW can efficiently and accurately describe quasi-particle correlated defect levels in two-dimensional (2D) monolayer MoS$_2$. By comparing two different defect states originating from sulfur vacancy and adatom to existing theoretical and experimental works, we show that our GW correction to the DFT defect levels is precisely modelled. Based on these findings, we expect that our method can provide genuine trap states for various 2D transition-metal dichalcogenide (TMD) monolayers, thus enabling the study of defect-induced effects on the device characteristics of these materials \textit{via} realistic simulations.
\end{abstract}


\section{Introduction}

The physical dimension of Si logic transistors is approaching the atomic limit, thus requiring novel architectures and/or high-mobility channel materials for future technology nodes. Logic switches based on two-dimensional (2D) transition-metal dichalcogenide (TMD) monolayers have thus been proposed to continue Moore's scaling law, thanks to their remarkable electronic properties. 
However, several works~\cite{RAI2022339,lee2019ab} reported that various defects inside these monolayers may limit their performance as logic devices, mainly through charged impurity scattering and defect-induced trap levels. In particular, the "mid-gap" states introduced by those impurities are presumably at the origin of large Schottky barriers (SB) and high contact resistances. 
Therefore, in order to understand the physics related to defects in 2D TMD monolayers and to guide device design, \textit{ab initio} simulations are required. 
In this work, we propose an efficient GW algorithm combined with density functional theory (DFT) to accurately describe defect levels in 2D TMD monolayers.
In conventional GW calculations, environmental effects from substrates are included to obtain the realistic bandgap of 2-D monolayers, which requires huge computational resources~\cite{naik2018substrate}. Our method, so-called projected GW (p-GW), overcomes this issue by projections onto a defect subspace while removing spurious interactions between periodic images by means of open boundary conditions. This algorithm can correctly predict the position of defect levels in the bandgap and ensure efficiency by resorting to the DFT-level bandgap.
We then apply this method to the most common defects in MoS$_2$ monolayers: S vacancy and adatom.

\section{Algorithm}

\begin{figure}[h]
     \centering
     \includegraphics[width=0.5\textwidth]{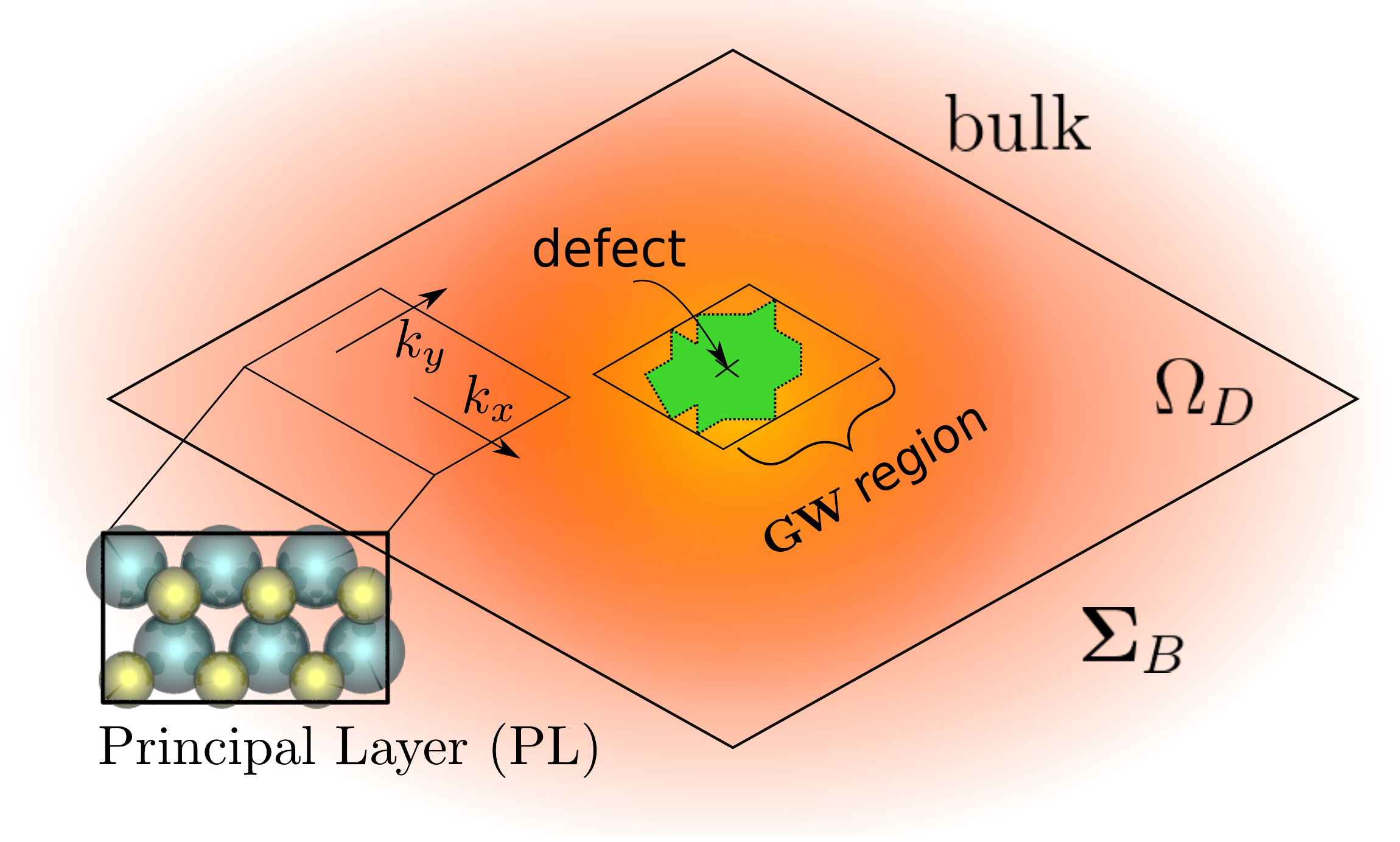}
     \caption{Schematic view of a device region $\Omega_D$ containing a defect and consisting of integer repetitions of a PL. The coupling to the bulk states of the surrounding material is described by $\mathbf{\Sigma}_B$. Electron-electron interactions beyond the mean-field level of theory are included only for a narrow region surrounding the defect and denoted as "GW region".}
     \label{fig:domain}
\end{figure}

\begin{figure*}[h]
    \centering
        \includegraphics[width=\textwidth, height=5.4cm]{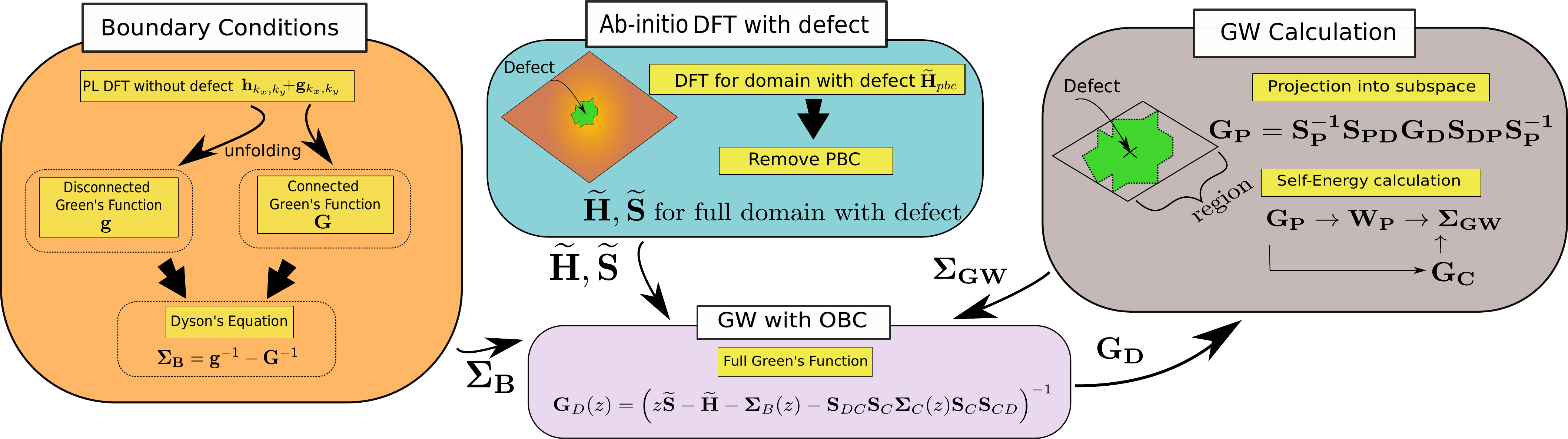}
              \caption{Flowchart of the p-GW method for efficient modelling of defected structures with GW corrections. (Orange box) $\mathbf{\Sigma}_B$ is constructed from the DFT calculation of a periodic PL. (Blue box) The $\mathbf{\widetilde{H}}$ and $\mathbf{\widetilde{S}}$ of the defected region are obtained from a separate DFT calculation by removing the PBCs. (Gray box) $\mathbf{\Sigma}_{GW}$ is computed for a subspace containing the defect where $\mathbf{G}_C$ is constrained to the defect and $\mathbf{W}_P$ includes its surroundings. (Purple box) Equation for the full Green's function coupling all boxes.}
            \label{fig:flowchart}
\end{figure*}

The p-GW algorithm is based on Green's function theory and aims at describing isolated defects.
The starting point is the definition of a device region $\Omega_D$ containing a defect and consisting of integer repetitions of a unit cell called "principal layer" (PL), as illustrated in Fig.~\ref{fig:domain}. 
We want to compute a device Green's function $\mathbf{G}_D$ which includes the correlation of the electrons localized around the defect and couples to the Bloch states of pristine MoS$_2$ at the boundaries.
This is achieved through the following procedure: 
\begin{enumerate}
\item A DFT calculation of $\Omega_D$ builds the Hamiltonian of the defect+MoS$_2$ system at a mean-field level.
\item A boundary self-energy replaces the periodic boundary conditions (PBCs) of DFT.
\item Projection onto an orthogonal subspace surrounding the defect defines the GW region.
\end{enumerate}
Our p-GW algorithm has been developed for the general case of a non-orthogonal basis set employed in the DFT calculation.
The flowchart in Fig.~\ref{fig:flowchart} shows how the above three steps can be used to obtain $\mathbf{G}_D$. We describe each step in more details below.

\subsection{Boundary self-energy} \label{subsec:sigmab}

The boundary self-energy $\mathbf{\Sigma}_B$ describes the coupling of $\Omega_D$ with the electrons in the Bloch states of MoS$_2$. $\Omega_D$ must be choosen large enough to safely assume that the potential at the boundaries is converged to the one of pristine MoS$_2$. $\mathbf{\Sigma}_B$ can then be efficiently obtained from a k-point calculation of the PL as summarized in the following (see also Refs.~\cite{gandus2020efficient,papior2019removing} for more details):
\begin{enumerate}
    \item An iterative method such as Sancho-Rubio~\cite{sancho1985highly} calculates the PL surface Green's function along one direction.
    \item The recursive Green's function algorithm expands to the size of the device region in the same direction.
    \item A partial Bloch's sum unfolds the Green's function in the remaining direction.
    \item The Dyson equation is used to compute $\mathbf{\Sigma}_B$.
\end{enumerate}
The last step requires the evaluation of a disconnected Green's function given by

\begin{equation}
    \label{eq:SigmaCorr}
    \mathbf{g}(z) = \left( z\mathbf{S} - \mathbf{H} \right)^{-1},
\end{equation}

where $z$ is a complex energy with an infinitesimal shift along the imaginary axis. $\mathbf{H}$ and $\mathbf{S}$ are the Hamiltonian and overlap matrices of the defect-free $\Omega_D$ region, which can be conveniently obtained from the corresponding PL matrices exploiting the translational symmetry of the supercell structure~\cite{papior2019removing}.
This efficient and precise algorithm allows us to treat the system as "open" and effectively simulates the defect as isolated. Indeed, this avoids undesired interferences or bound state patterns related to the PBCs.

\subsection{Projection onto a GW region}

GW corrections are computed only for a narrow region $\Omega_C$ surrounding the defect, as shown in Fig.~\ref{fig:domain}. We start by selecting a region $\Omega_P \subset \Omega_D$ where the formation of electron-hole pairs is expected to screen the electrons in the defect level states. Because of the strong atomic orbital character of these states, this region can only comprise the defect and a few atoms nearby. Taking into account the non-orthogonality between the DFT basis set, we write the Green's function projected onto $\Omega_P$ as \cite{jacob2015towards}

\begin{equation}
\label{eq:projectedGreen}
    \mathbf{G}_P = \mathbf{S}_{P}^{-1}\mathbf{S}_{PD}\mathbf{G}_{D}\mathbf{S}_{DP}\mathbf{S}_{P}^{-1},
\end{equation}

where $\mathbf{S}_{DP}$ is the overlap matrix between orbitals in $\Omega_D$ and in $\Omega_P$, respectively, and $\mathbf{S}_{P}$ is the overlap matrix within $\Omega_P$. Starting from $\mathbf{G}_P$ and the bare Coulomb matrix calculated for the basis functions in $\Omega_P$, we obtain the screened Coulomb interaction $\mathbf{W}_P$ within the random phase approximation~\cite{strangeGW,rostgaard2010fully}. Finally, we multiply the part of the screened interaction $\mathbf{W}_C$ in $\Omega_C$ by the Green's function $\mathbf{G}_C$ projected from $\mathbf{G}_P$ in analogy with Eq.~\ref{eq:projectedGreen} to obtain the GW self-energy $\mathbf{\Sigma}_{GW}$.


\subsection{Device Green's function}

The device Green's function allows to access the electronic and transport properties of the defected monolayers. It is obtained as\cite{thygesen2008conserving}

\begin{equation}
    \label{eq:FullGreen}
    \mathbf{G}_D(z) = \left( z \widetilde{\mathbf{S}} - \widetilde{\mathbf{H}} - \mathbf{\Sigma}_{B}(z) - \mathbf{S}_{DC}\mathbf{S}_{C}\mathbf{\Sigma}_C(z)\mathbf{S}_{C}\mathbf{S}_{CD} \right)^{-1},
\end{equation}

where $\widetilde{\mathbf{H}}$ and  $\widetilde{\mathbf{S}}$ are the device Hamiltonian and overlap matrices with removed PBCs and

\begin{equation}
    \label{eq:SigmaCorr}
    \mathbf{\Sigma}_C(z) = - \mathbf{V}_{xc} + \mathbf{\Sigma}_{GW}(z) + \delta \mathbf{V}_{H},
\end{equation}

where $\mathbf{V}_{xc}$ is the DFT exchange-correlation (XC) potential that needs to be subtracted to avoid double counting of the correlations included in $\mathbf{\Sigma}_{GW}$. $\delta \mathbf{V}_{H}$ is the deviation from the DFT Hartree potential and is calculated from the change in the density matrix $\mathbf{D}_C$ in the $\Omega_C$ region. Because $\delta \mathbf{V}_{H}$ and $\mathbf{\Sigma}_{GW}$ depend on $\mathbf{G}_D$ themselves, Eq.~\ref{eq:FullGreen} is solved self-consistently until convergence of $\mathbf{D}_C$.

\section{Results}
\begin{figure}[h]
     \centering
     \begin{subfigure}[t]{0.33\textwidth}
         \centering
         \includegraphics[width=\textwidth]{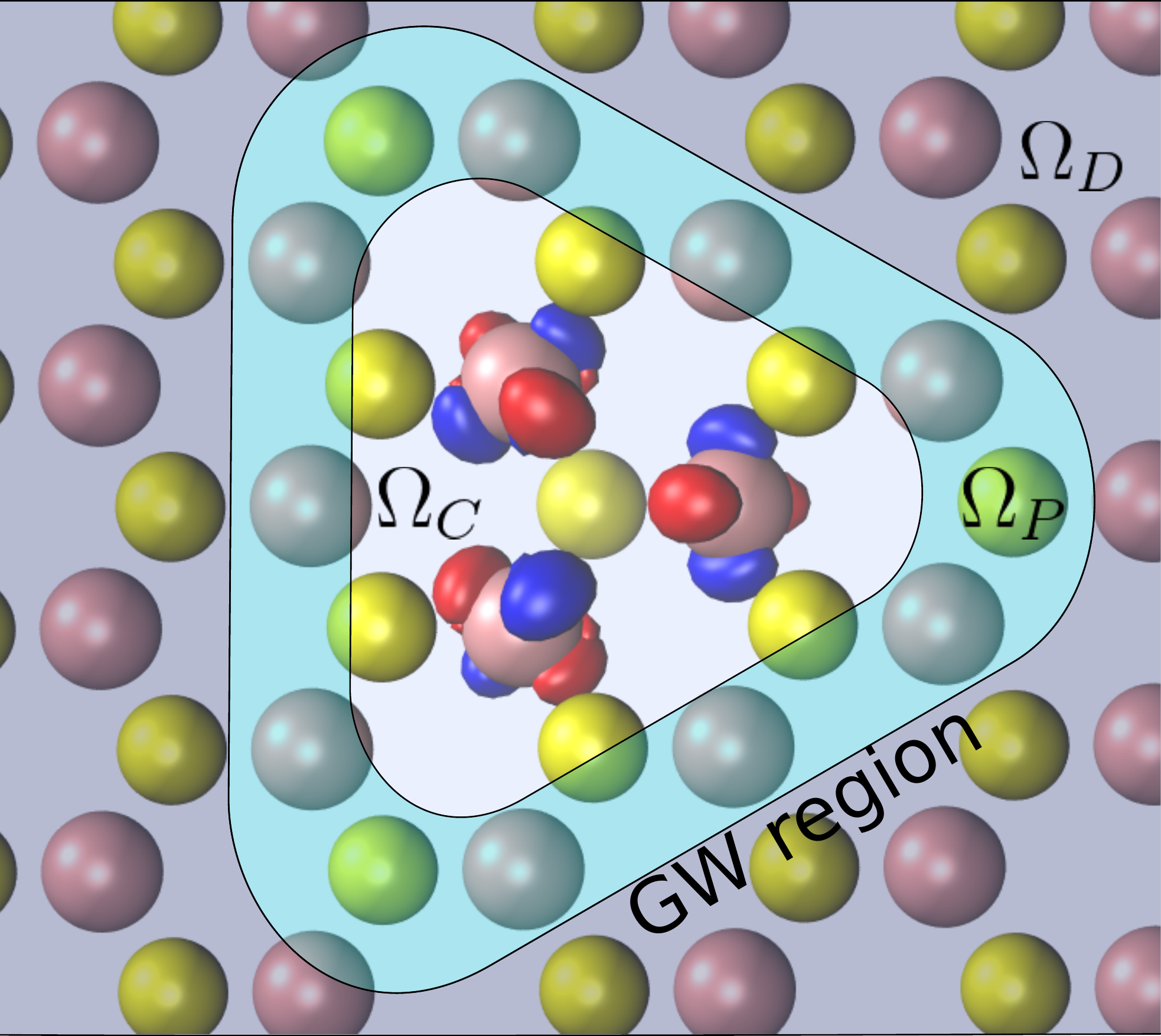}
         \caption{S-.}
         \label{fig:domainSminus}
     \end{subfigure}%
     \hspace{2em}%
     \begin{subfigure}[t]{0.33\textwidth}
         \centering
         \includegraphics[width=\textwidth]{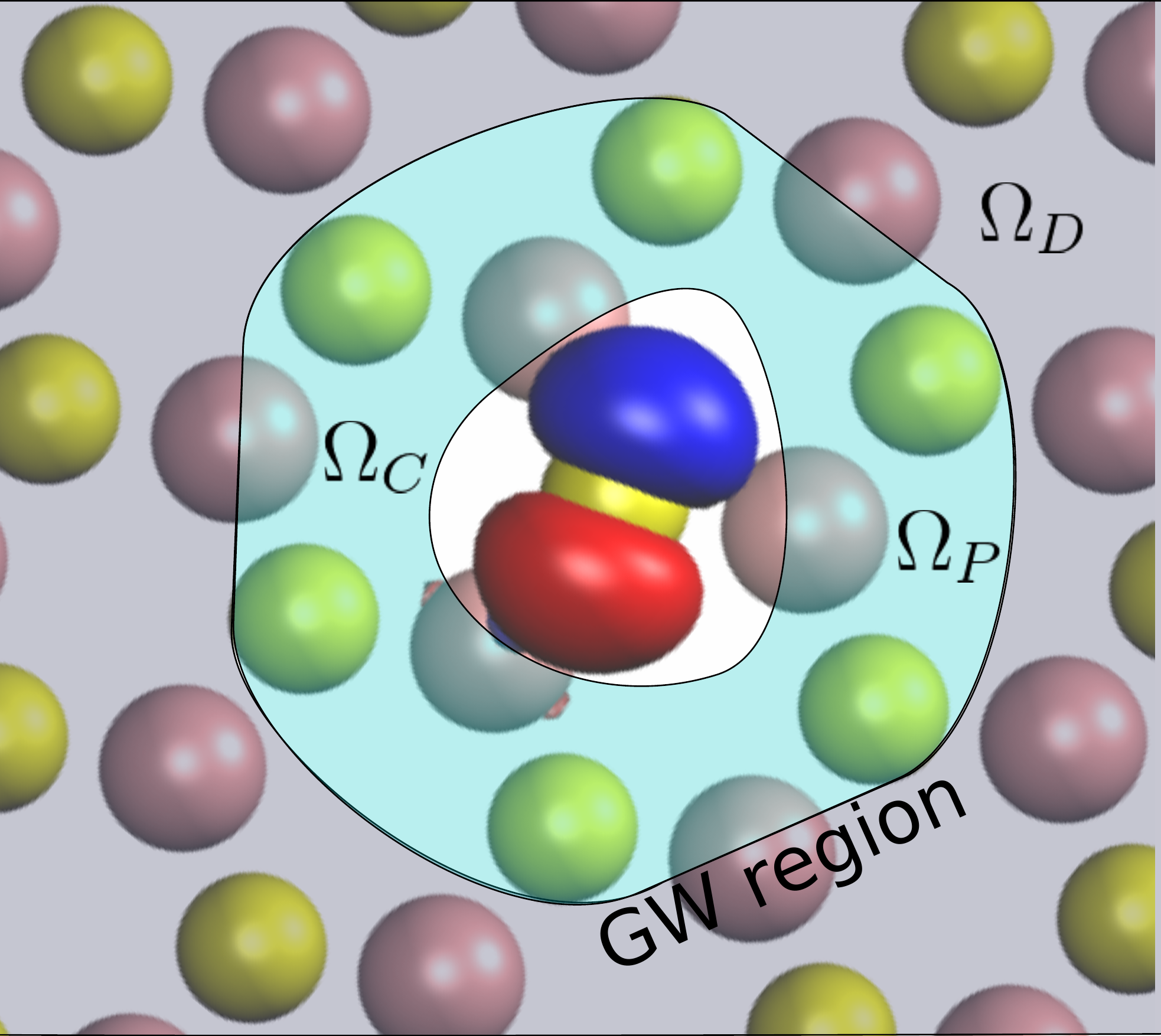}
         \caption{S+.}
         \label{fig:domainSplus}
     \end{subfigure}%
     \caption{GW region considered in this work to model S vacancies (a) and adatoms (b) in MoS$_2$ monolayers. The isosurfaces represent the wavefunction of the state created by the defect. For S+, there's an additional degenerate defect state with an electron in the $p$ orbital orthogonal to the one shown here.}
     \label{fig:domainS}
\end{figure}

We study the effect of S vacancies (S-) and adatoms (S+) in 2D MoS$_2$ monolayers.
The device region is composed of $4\times6$ repetitions of a PL composed of $6$ Mo and $12$ S atoms, as shown in Fig.~\ref{fig:domain}.
The electronic structure calculation of the PL is over-sampled with a $11\times6\times1$ k-mesh to obtain a $\mathbf{\Sigma}_B$ that precisely describes the bulk MoS$_2$ states. 
The GW region is shown in Figs.~\ref{fig:domainSminus} and \ref{fig:domainSplus} together with the wavefunction of the states created by the defect. $\Omega_P$ includes up to the $2$nd nearest neighbor to the defect, i.e. $12$ Mo and $13$ S atoms for S- and $3$ Mo and $15$ S atoms for S+. The defect states have a strong atomic orbital character: they are essentially a superposition of the $3d$ Mo orbitals closest to the vacancy for S- and an unpaired electron in the in-plane $p$ orbital of the S adatom for S+. This allows us to define $\Omega_C$ as the $3$ Mo and their surrounding S atoms for S- and the single S adatom for S+. $\mathbf{\Sigma}_C$ is then computed in this region only.

\begin{figure*}[h]
     \centering
     \begin{subfigure}[t]{0.33\linewidth}
         \centering
         \includegraphics[width=\textwidth]{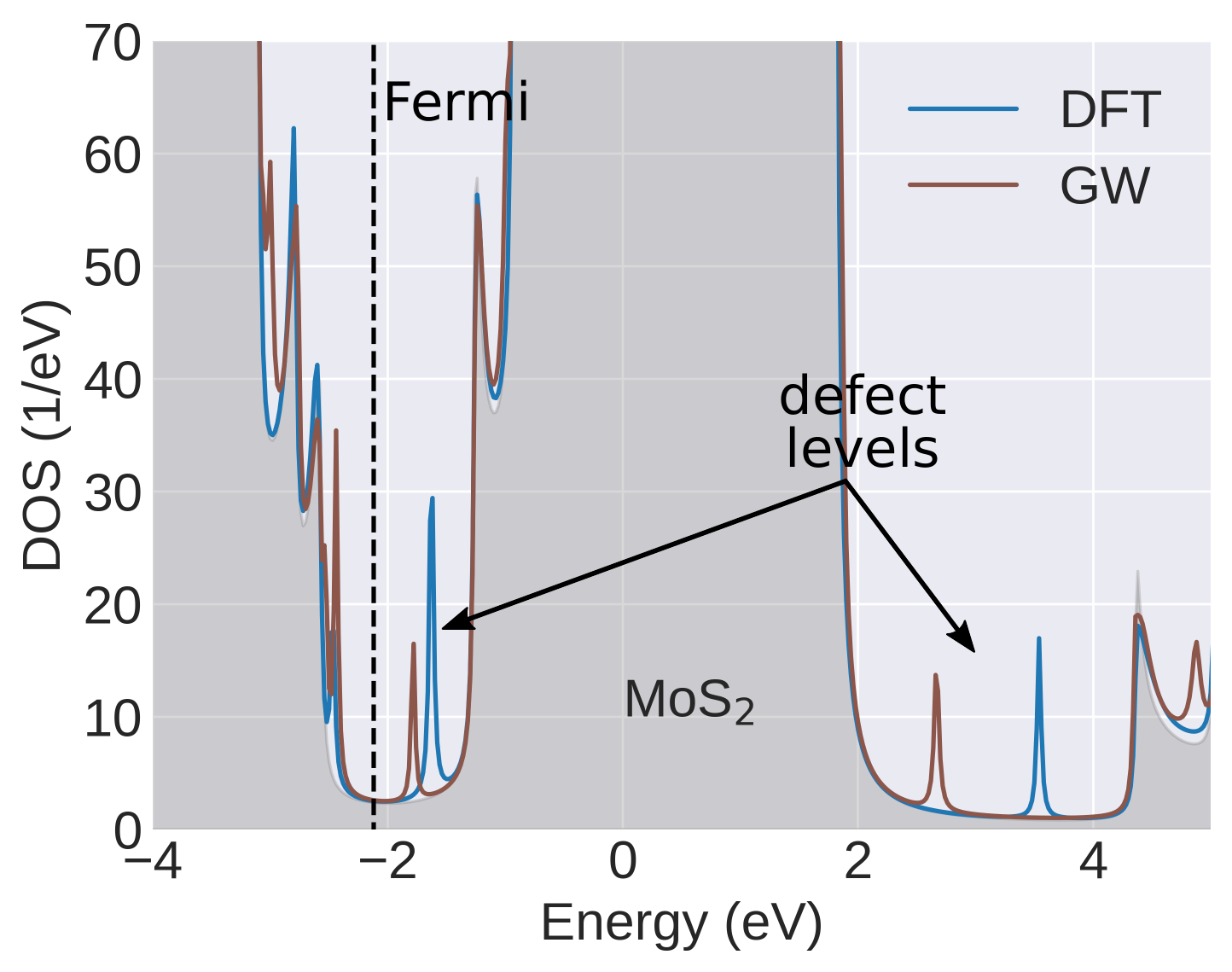}
         \caption{Density-of-states.}
         \label{fig:dos}
     \end{subfigure}%
     \hspace{0em}%
     \begin{subfigure}[t]{0.33\linewidth}
         \centering
         \includegraphics[width=\textwidth]{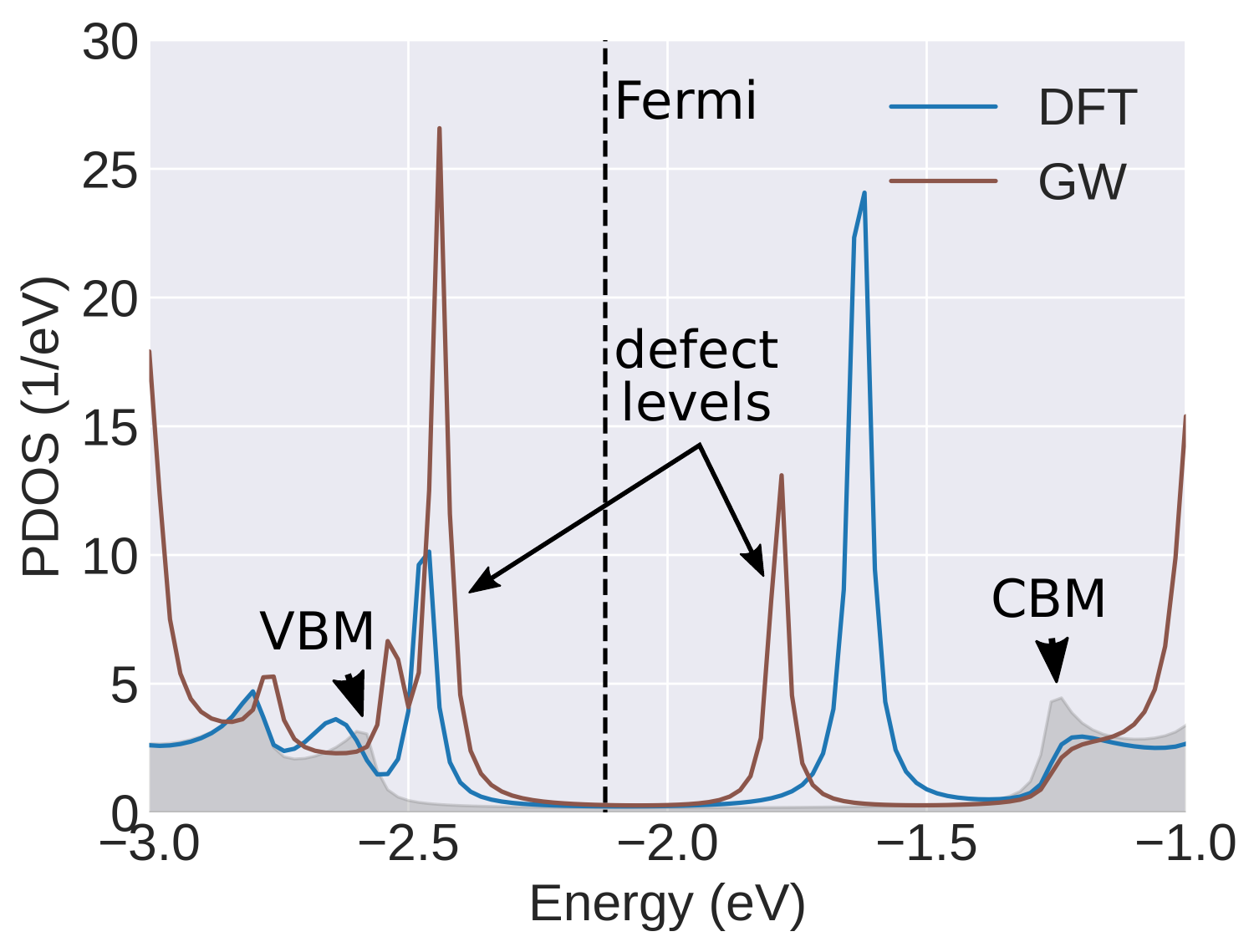}
         \caption{Projected density-of-states.}
         \label{fig:pdos}
     \end{subfigure}%
     \hspace{0em}%
     \begin{subfigure}[t]{0.33\linewidth}
         \centering
         \includegraphics[width=\textwidth]{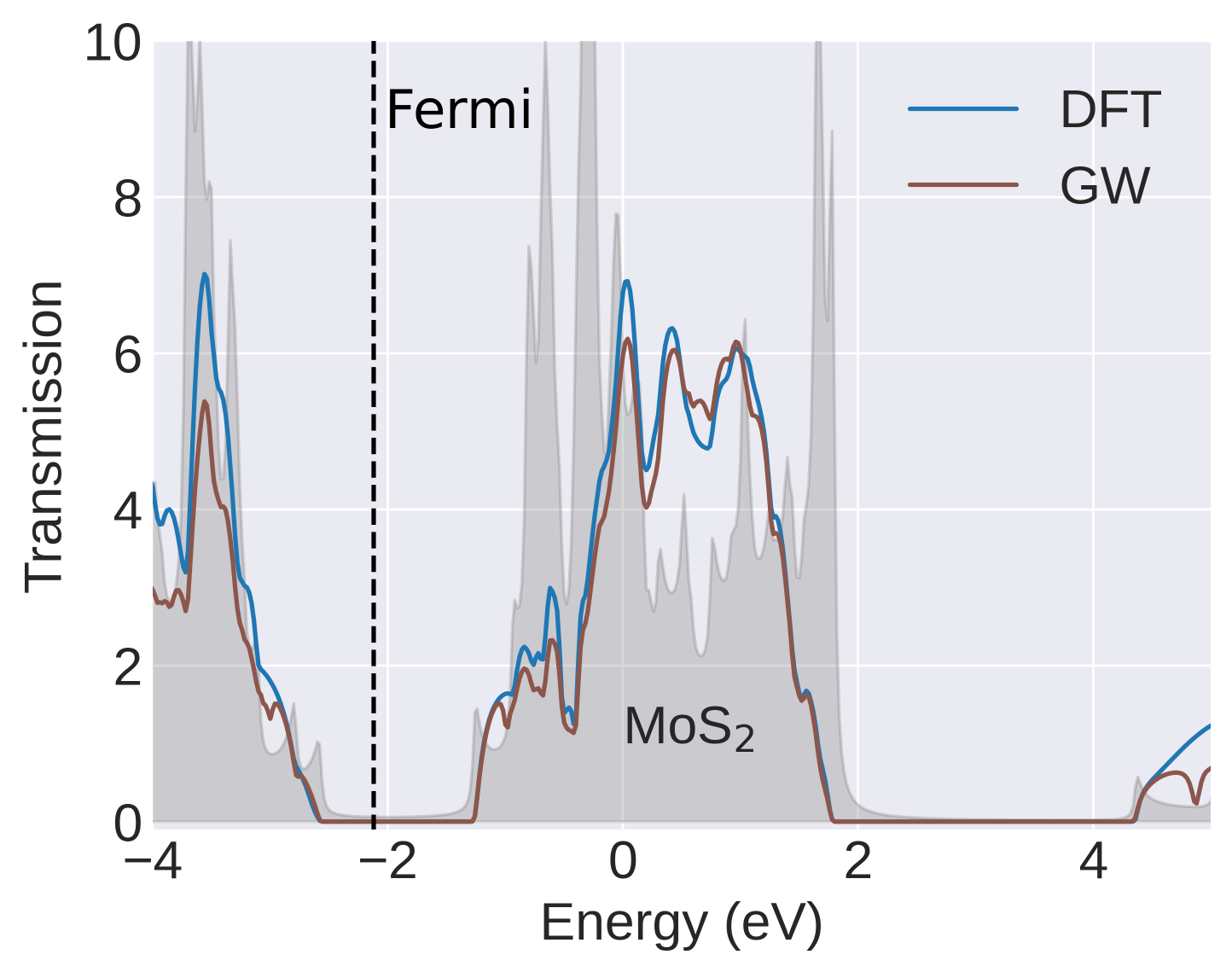}
         \caption{Transmission function.}
         \label{fig:trans}
     \end{subfigure}
     \centering
     \begin{subfigure}[t]{0.33\linewidth}
         \centering
         \includegraphics[width=\textwidth]{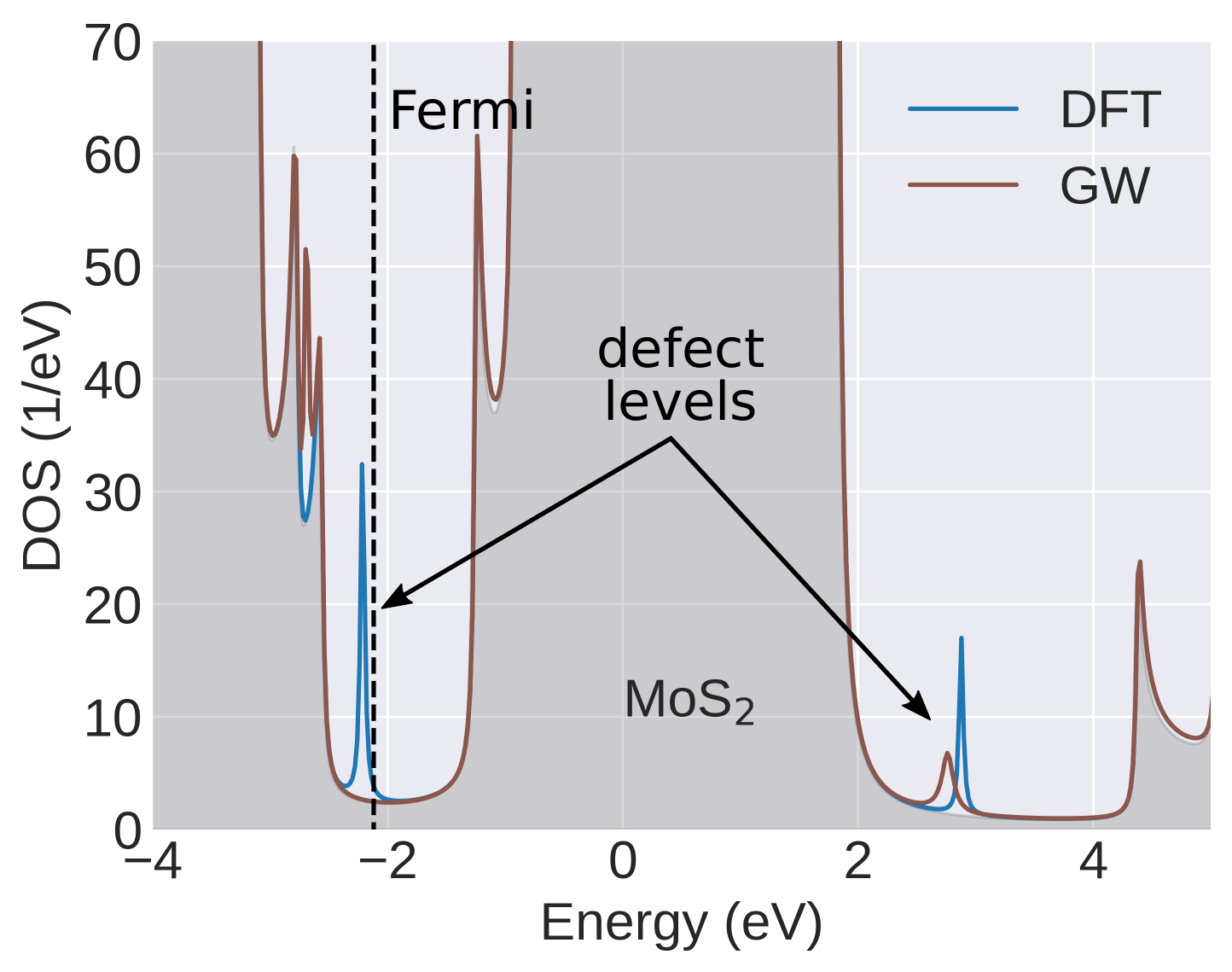}
         \caption{Density-of-states.}
         \label{fig:dos}
     \end{subfigure}%
     \hspace{0em}%
     \begin{subfigure}[t]{0.33\linewidth}
         \centering
         \includegraphics[width=\textwidth]{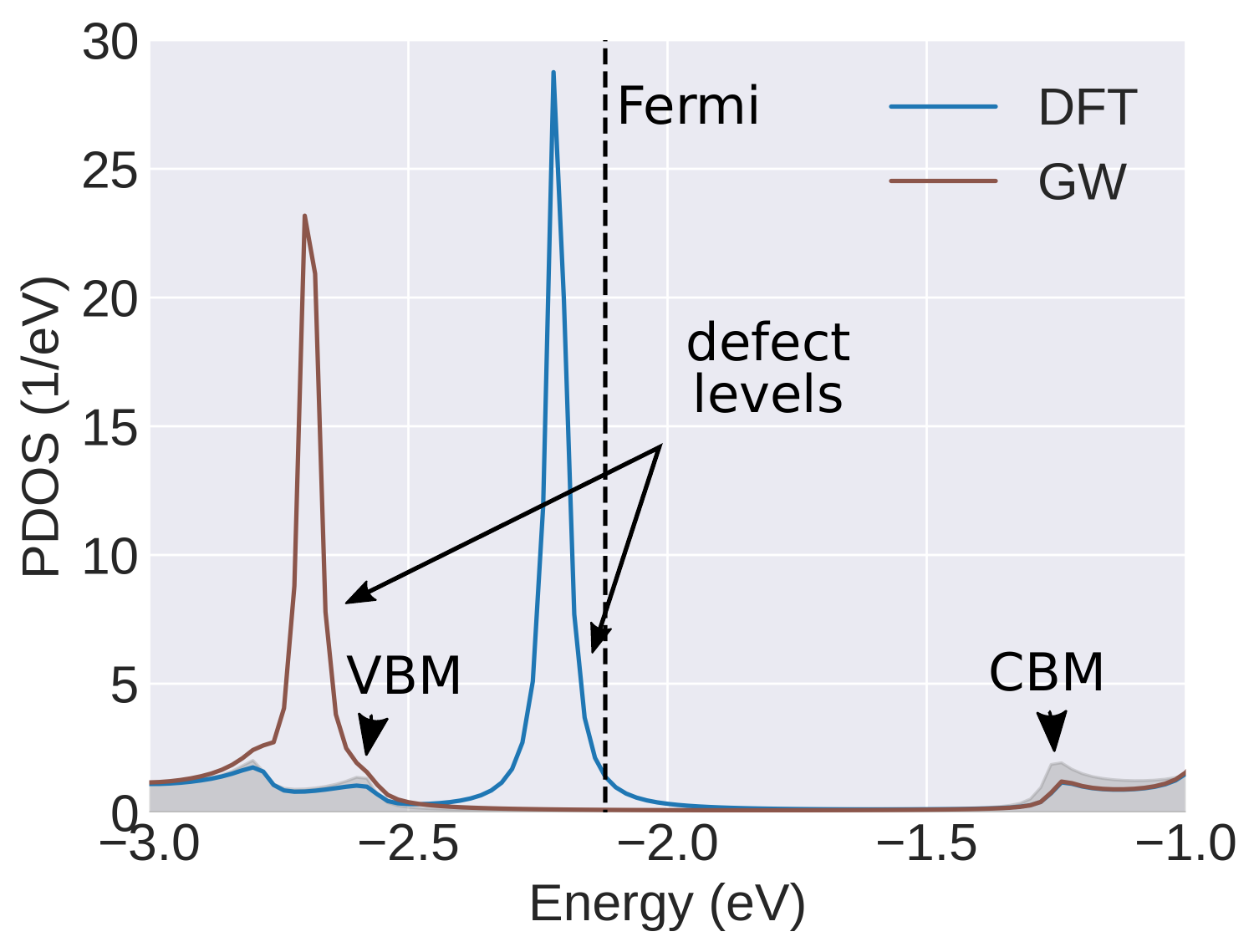}
         \caption{Projected density-of-states.}
         \label{fig:pdos}
     \end{subfigure}%
     \hspace{0em}%
     \begin{subfigure}[t]{0.33\linewidth}
         \centering
         \includegraphics[width=\textwidth]{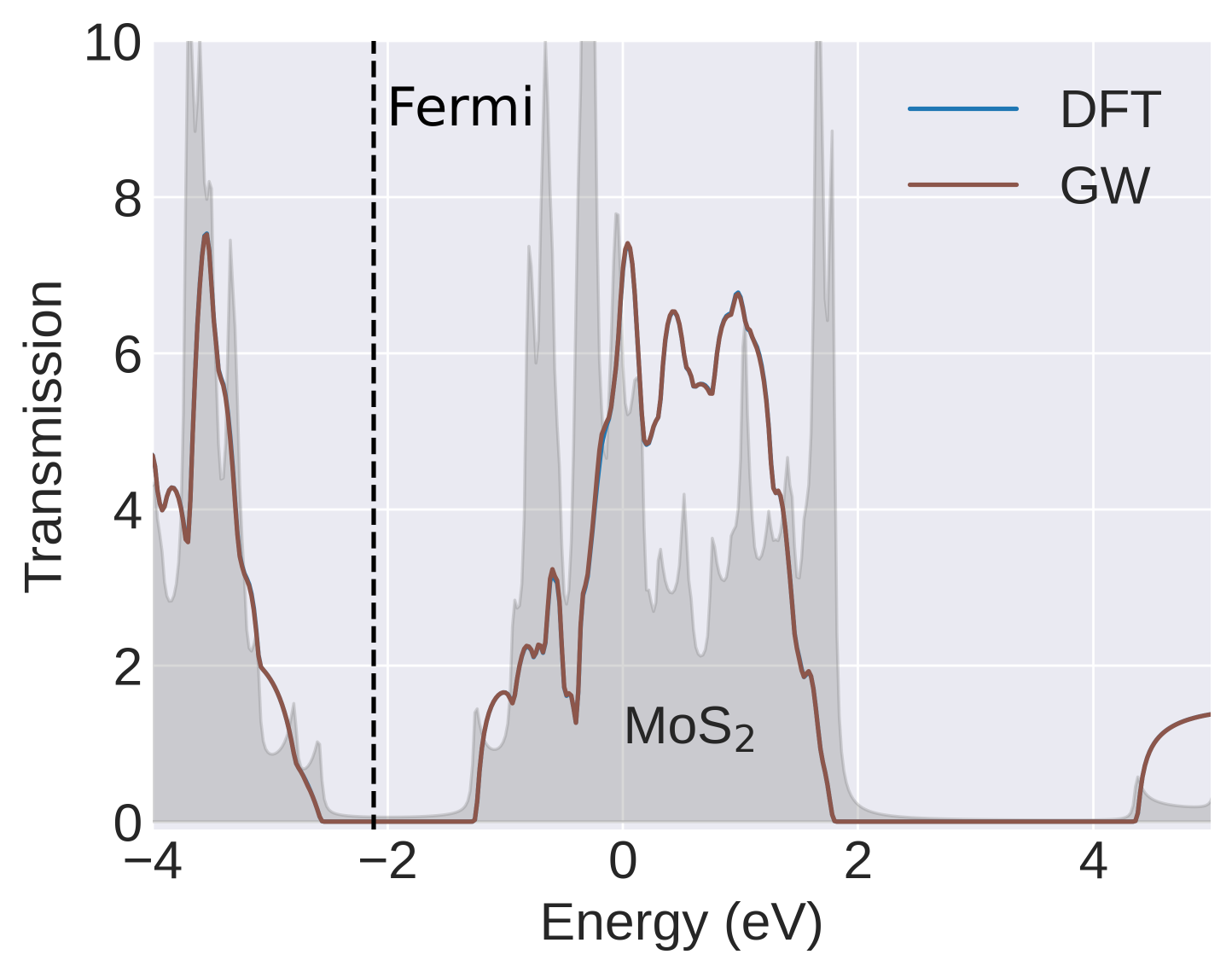}
         \caption{Transmission function.}
         \label{fig:trans}
     \end{subfigure}
        \caption{Results for S- (a-c) and S+ (d-f) obtained by DFT and the proposed p-GW method. The total density-of-states of the device region (a,d) and projected onto the GW subspace (b,e) show that the many-body correction shifts the defect level states while maintaining the DFT bulk properties. (c,f) Transmission function through the defected structure. The onset of the electron transmission around the fundamental gap is also preserved by the p-GW correction.}
        \label{fig:results}
\end{figure*}

We calculated the correspondingdensity-of-states (DOS), the projected DOS (PDOS), and the electron transmission and report these results in Fig.~\ref{fig:results}. It is apparent from the DOS and the PDOS that the effect of the many-body correction is to shift the energy levels of the defect while preserving the DFT properties, i.e. the bandgap, as also corroborated by the conservation of the bulk-like electronic transmission. Previous k-point GW studies of full defect+MoS$_2$ S- structures found similar positions of the defect level with respect to the corresponding band edges~\cite{naik2018substrate}.
This indicates that our p-GW algorithm can accurately predict trap-levels with minimal computational burden.
The DFT study for S+ predicts a shallow state close to the valence band. The GW correction pulls the defect-level position down into the valence band, as suggested by experimental studies that show a strong p-type behaviour in the presence of S adatoms \cite{addou2015surface}.  
This indicates that such defects may act as doping centre, a behaviour that is not captured by pure DFT.

\section{Conclusions} 
We proposed a novel algorithm to locally and efficiently apply many-body corrections using GW to a region surrounding a defect. Periodic self-interactions are removed by virtue of an efficient boundary self-energy calculation. The presented algorithm is then applied to S vacancy and adatom defects in a MoS$_2$ monolayer. 
Our method is a first step toward the inclusion of many-body methods beyond DFT in large scale simulations of realistic devices.

\section*{Acknowledgment}
This work was supported by the NCCR MARVEL funded by the Swiss National Science Foundation grant 51NF40-205602. Computational support from the Swiss Supercomputing Center (CSCS) under project ID s1119 is gratefully acknowledged

\bibliographystyle{unsrt}  
\bibliography{refs}

\end{document}